\documentclass[12pt,amsfonts]{article}
\usepackage{amsfonts}
\def\one{1\hskip-.37em 1}

\def\mfH{\mathfrak{H}}

\def\ra{\rightarrow}
\def\tint{{\textstyle\int}}

\def\s{\hskip.08em}

\def\d{\partial}

\def\b{\begin{eqnarray*}}  
\def\e{\end{eqnarray*}}    
\def\bn{\begin{eqnarray}}  
\def\en{\end{eqnarray}}   
\def\ra{\longrightarrow}
\def\<{\langle}
\def\>{\rangle}

\def\no{\nonumber}

\def\{{\lbrace}
\def\}{\rbrace}
\title{When Canonical Quantization Fails,\\ Here is How to Fix It}
\author{John R. Klauder\footnote{Email: john.klauder@gmail.com}\\
Department of Physics and Department of Mathematics\\
University of Florida,
Gainesville, FL 32611-8440}
\begin{document}
\maketitle
\begin{abstract} Following Dirac \cite{dirac}, the rules of canonical quantization include classical and quantum contact transformations of classical
and quantum phase space variables. While arbitrary classical canonical coordinate transformations exist that is not the case for some analogous
quantum canonical coordinate transformations. This failure is due to the rigid connection of quantum variables arising by promoting the corresponding
classical variable from a $c$-number to a $q$-number. A different relationship of $c$-numbers and $q$-numbers in the procedures of Enhanced
Quantization \cite{kla1} shows the compatibility of all quantum operators with all classical canonical coordinate transformations.
\end{abstract}
\section{Canonical Quantization: A Mini-review}
 For a single degree of freedom, the classical phase-space variables $p$ and $q$ are real and have a standard Poisson bracket  $\{q,p\}=1$. They are
 used to define the real classical Hamiltonian $H(p,q)$, which determines the dynamics by Hamilton's equations of motion $\dot{q}(t)=\d H(p,q)/\d p(t)$
 and $\dot{p}(t)=-\d H(p,q)
/\d q(t)$. Contact transformations (a.k.a.~canonical transformations) lead to the introduction of new, real canonical coordinates $q^*=q^*(p,q)$ and
$p^*=p^*(p,q)$ where $p^*\s dq^*=p\s\s dq+dF(q^*,q)$ (or a straightforward variation of this equation), which leads to $\{q^*,p^*\}=1$ and
 $H^*(p^*,q^*)=H(p,q)$. Dynamics is then given by $\dot{q}^*(t)=\d H^*(p^*,q^*)/\d p^*(t)$ and $\dot{p}^*(t)=-\d H^*(p^*,q^*)/\d q^*(t)$.

Corresponding canonical quantum operators are given by `promoting' classical canonical variables to Hermitian operators acting on vectors in an
infinite-dimensional Hilbert space. This leads to connections such as $p\!\ra\! P$ and $q\!\ra\! Q$, where $[Q,P]\equiv QP-PQ=i\hbar\one$, and
$\hbar=h/2\pi$ is the
reduced Planck's constant. In
addition, the classical Hamiltonian $H(p,q)\!\ra\!{\cal H}(P,Q)$, and {\it in the right coordinate system}, ${\cal H}(P,Q)=H(P,Q)+{\cal O}(\hbar;P,Q)$;
namely, in the right coordinates, the Hamiltonian operator is the same function of $P$ and $Q$ as the classical Hamiltonian is of $p$ and $q$, apart from a
contribution that involves some possible $\hbar$ corrections \cite{dirac} (page114).
 The `right coordinate system' asserts that the classical coordinates must be Cartesian coordinates.
Phase space is not normally equipped with a metric and thus a metric must be added to the usual phase space. A Cartesian behavior
is plausible for higher-dimensional systems, e.g., two dimensions, so one can assume a flat coordinate space $dq_1^2+dq_2^2$, with
$-\infty<q_1,q_2<\infty$,  and a flat momentum space $dp_1^2+dp_2^2$, with $-\infty<p_1,p_2<\infty$, and this is what most texts implicitly
assume. However, that choice does not automatically fit a {\it one}-dimensional phase space with a single $q$ and a single $p$ since, by itself,
one dimension cannot be Cartesian; however, one could imagine there are two, separate, identical, independent, one-dimensional dynamical systems,
and then just focus on one of them. Of course, the need to consider Cartesian coordinates is necessary since the spectrum of the quantum Hamiltonian
operator specifies the energy levels and in most systems the energy levels are observable quantities.
While normally there are classical coordinates such that $H^*(p^*,q^*)=p^*$ one should not choose ${\cal H}^*(P^*,Q^*)=P^*$ since the spectrum of
$P^*$ would most likely be in the continuum, e.g., a continuous spectrum from $0$ to $\infty$, which would generally differ from the proper spectrum
offered by Cartesian coordinates.
Thus, as decades of successful applications confirm, the special role played by Cartesian coordinates in canonical quantization is essential.

\section{Classical and Quantum \\Contact Transformations}
On page 105 of \cite{dirac} Dirac establishes his interpretation of all contact transformations in which the classical transformation of
$(p,q)\!\ra\!(p^*,q^*)$ (Dirac's notation) has a role to play in the quantum theory. While $q$ and $q^*$ are real and $Q$ and $Q^*$ are Hermitian (for 
terminology see Appendix),
that does not ensure the operators are self adjoint (see Appendix). To illustrate the point, we note that the Schr\"odinger representation
for $Q$ is $x$,
where $-\infty<x<\infty$, and for $P$ is $-i\hbar\s\d/\d x$. Equivalently, one can choose $Q$ as $i\hbar\s\d/\d k$ and $P$ as $k$, where
$-\infty<k<\infty$. In these variables we have $\one=\tint |x\>\<x|\,dx=\tint|k\>\<k|\,dk$ (undefined integration limits are from $-\infty$
to $+\infty$) where $Q|x\>=x|x\>$, $\<x|x'\>=\delta(x-x')$,  and $P|k\>=k|k\>$,
$\<k|k'\>=\delta(k-k')$, and $\psi(x)\equiv\<x|\psi\>$ while
$\widetilde{\psi}(k)\equiv\<k|\psi\>$ for all abstract vectors $|\psi\>\in\mfH$. It follows that
$\<x|k\>= (2\pi\hbar)^{-1/2}\,\exp[ikx/\hbar]=\overline{\<k|x\>}$; the over-line denotes complex conjugation.

However, for some dynamical systems  the classical coordinate $q$ satisfies $q>0$, which implies that $Q>0$ as well. In such
a case $Q|x\>=x|x\>$, $\<x|x'\>=\delta(x-x')$ still hold along with $\one=\tint_0^\infty|x\>\<x|\,dx$. However, the operator $P$
now has some imaginary eigenvalues as is clear from the example $P\psi(x)=-i\hbar\d\psi(x)/\d x=i\alpha\s\psi(x)$, $0<\alpha<\infty$, with
solutions given by $\psi(x)=\exp[-\alpha x/\hbar]$ that are
genuine elements in the Hilbert space. In this case, $P$ is Hermitian but not self adjoint and it can not be made self adjoint (see the
Appendix for a definition of these terms). Thus, while the classical variables $p$ and $q>0$ are perfectly acceptable for some systems,
the corresponding operators $P$ and $Q>0$ do not represent observables since $P$ is ill behaved. A simple contact transformation leads to
$p^*=-q<0$ and $q^*=p$, which means that  the quantum operator $Q^*$ now is ill behaved.  Moreover, while the contact transformation of the classical variables $p$ and $q>0$ given by $p^*=\sqrt{2}\s p+q$ and $q^*=\sqrt{2}\s q+p$ is a valid classical contact transformation, the quantum version of these variables shows that neither $P^*=\sqrt{2}\s P+Q$  nor $Q^*=\sqrt{2}\s Q+P$ are self adjoint and thus {\it both operators} are unsuitable as quantum observables. (This kind of difficulty applies even if $q>-\gamma$ where $\gamma>0$, e.g., $\gamma=10^{100000}$!)

Dirac used the presumed validity of the quantum version of variables such as $P^*$ and $Q^*$ in discussing how quantum contact transformations
form a subset of unitary transformations; see Eq.~(77) on page 106 of \cite{dirac} where he argues that contact transformations
can be seen as unitary transformations. The examples discussed in the preceding paragraph show that sometimes this assumption is false.

\section{Enhanced Quantization}
This section illustrates a few features in a new procedure to relate classical and quantum systems to one another; a full account
of all the features in this new approach is available in \cite{kla1} and references therein.
Enhanced quantization is designed to ensure that the classical formalism is a `natural subset' of the quantum formalism since in the real
world $\hbar$ {\it is not zero}.
The main idea in this alternative quantization procedure is how classical and quantum variables relate to each other. The quantum action
principle is given by $A_Q=\tint_0^T\s\<\psi(t)|\s[i\hbar\s \d/\d t-{\cal{H}}(P,Q)\s]\s|\psi(t)\>\,dt$, and general variations of
$\{|\psi(t)\>\}_0^T$, holding the end points $|\psi(0)\>$ and $|\psi(T)\>$ fixed, leads to Schr\"odinger's equation
$i\hbar\s\d\s|\psi(t)\>/\d t={\cal{H}}(P,Q)\s|\psi(t)\>$. However, classical observers can only vary the position and velocity
of such a system, and, thanks to Galilean invariance, these changes can be made by the observer alone and thus the quantum system is not
disturbed. Choosing a normalized fiducial vector $|0\>$, which, for example, satisfies $(Q+iP)\,|0\>=0$, means that we can only vary
normalized vectors of the form $|p,q\>\equiv \exp[-iqP/\hbar]\s\exp[ipQ/\hbar]\s|0\>$, thanks to the relation $\dot{q}=\d H(p,q)/\d p(t)$;
these vectors for all $(p,q)\in
{\bf R}^2$ are also known as {\it canonical coherent states} \cite{BSJK} and the operators $P$ and $Q$ must both be self adjoint since
only such operators can generate unitary, one-parameter groups. This vector limitation means that a classical observer may seek
(R for restricted) variations of
\bn  A_{Q(R)}&&\hskip-1.2em=\int_0^T\<p(t),q(t)|\s[\s i\hbar\s \d/\d t-{\cal{H}}(P,Q)\s]\s|p(t),q(t)\>\,dt\no\\
    &&\hskip-1.22em=\int_0^T\s[\s p(t)\s \dot{q}(t)-H(p(t),q(t))\s]\,dt\;, \label{e1} \en
where, assuming that ${\cal H}(P,Q)$ is a polynomial for illustration,
\bn  H(p,q)\hskip-1.3em&&=\<p,q|\s{\cal H}(P,Q)\s|p,q\>\no\\
&&=\<0|\s{\cal H}(P+p\one,Q+q\one)\s|0\> \label{e2} \\
&&={\cal H}(p,q)+{\cal O}(\hbar;p,q)\;.\no  \en
The form of (\ref{e1}), along with (\ref{e2}), leads to the {\it enhanced classical action functional} thanks to the fact that $\hbar$
has not changed from its positive, physical value and it may play an important role in solutions to the equations of motion.
Note well: The functional form of the Hamiltonian operator $({\cal H})$ here is {\it exactly} that of the classical Hamiltonian function $(H)$
apart from possible order-$\hbar$ corrections, just as the Cartesian rule requires. Phase space may not have a metric but Hilbert space has one,
and the distance between two infinitely close coherent states,
 ignoring any phase difference, is determined by an appropriate (suitably scaled) Fubini-Study metric \cite{w} given by
$ds^2\equiv (2\hbar)\s[\s\|\s d|p,q\>\|^2-|\<p,q|\s d|p,q\>|^2\s]$,
and which, for our example, leads to $ds^2=dp^2+dq^2$, a Cartesian metric all its own! This nice expression is partly due to our choice
of the fiducial vector $|0\>$ which obeys $(Q+iP)\s|0\>=0$, but the result for a general normalized fiducial vector, say $|\eta\>$, is given by
 \bn    ds^2=(2/\hbar)\s\{\s dp^2\,\<(\Delta Q)^2\>+dp\s dq\s\<[\s\Delta Q\s\Delta P+\Delta P\s\Delta Q\s]\>+dq^2\,\<(\Delta P)^2\>\s\}\;, \label{e3}  \en
 where $\<(\cdot)\>\equiv \<\eta|(\cdot)|\eta\>$ and $\Delta X\equiv X-\<X\>$. Clearly, a suitable linear shift of the coordinates
 immediately leads to a Cartesian metric even for a general choice of $|\eta\>$.

 Note that $p$ and $q$ in (\ref{e1}) and (\ref{e2}) are arbitrary and independent real numbers, and, along with the fact that $\hbar>0$
 as well, this amounts
 to an enhanced classical variational principle as part of the quantum variational principle. Thus we have achieved our main goal of showing
 that the classical story is part of the quantum story.
 Moreover, since $(p,q)$ are completely independent of $(P,Q)$ we are free to make any contact transformation we like, e.g. $(p,q)\!\ra\! (p^*,q^*)$.
 In so doing, the vector
   \bn |p,q\>\!\ra\! |p^*,q^*\>\equiv e^{-iq(p^*,q^*)\s P/\hbar}\,e^{ip(p^*,q^*)\s Q/\hbar}\,|0\>\;. \label{e4}\en
But, after all, the label $(p,q)$ served to identify a specific vector in $\mfH$, and the new label $(p^*,q^*)$ is supposed to identify
 {\it the very same   vector}, and therefore $|p^*,q^*\>\equiv|p(p^*,q^*), q(p^*,q^*)\>=|p,q\>$. Along with the relation
 $p\s\s dq= p^*\s dq^* + dG^*(p^*,q^*)$ for a suitable generator $G^*$, it follows that in the new coordinates
  \bn  A_{Q(R)}\hskip-1,3em&&=\int_0^T\s\<p^*(t),q^*(t)|\s[\s i\hbar\s\d/\d t-{\cal H}(P,Q)\s]\s|p^*(t),q^*(t)\>\,dt \no\\
         &&=\int_0^T [\s p^*(t)\s\dot{q}^*(t)+{\dot G}^*(p^*(t),q^*(t))-H^*(p^*(t),q^*(t))\s]\,dt\;, \en
 which is the correct enhanced classical action functional expressed in the new variables {\it without any modification of the quantum operators whatsoever!}

 \subsection{Affine variables}
 The previous discussion dealt with classical systems for which both `Cartesian' phase-space variables $p$ and $q$ range over the entire real line.
 However, that study can not deal with situations where $q>0$ (or $q>-\gamma$), and to handle such situations we need to use new
 kinematical operators. Although $P$ is ill behaved when $Q>0$, the dilation operator $D\equiv (PQ+QP)/2$ is self adjoint and along
 with $Q>0$ (or $Q<0$) the operators $(D,Q)$ denote
 {\it affine variables} that satisfy
  the Lie algebra $[Q,D]=i\hbar\s Q$. While there are two, principal, self-adjoint realizations, one where $Q>0$ and the other
  where $Q<0$, we will focus on the
 case where $Q>0$. Choosing $Q$ to be dimension free, we select a new, normalized fiducial vector, $|\beta\>$, chosen as a
 solution of the equation $[\s (Q-1)+iD/\beta\hbar\s]\s|\beta\>=0$,
 which ensures that $\<\beta|Q|\beta\>=1$ and $\<\beta|D|\beta\>=0$. The quantum action functional is the same as before except
 that now the Hamiltonian operator ${\cal{H}}'={\cal H}'(D,Q)$. A classical observer can only vary $p$ and $q>0$ as arguments in
 the {\it affine coherent states} $|p,q;\beta \>\equiv
 \exp[ip\s Q/\hbar]\s\exp[-i\ln(q) D/\hbar]\,|\beta\>$ \cite{kla2,BSJK}. This restriction leads to the enhanced classical action functional
  \bn A_{Q(R)}\hskip-1.2em&&=\int_0^T\<p(t),q(t);\beta|\s[\s i\hbar\s\d/\d t-{\cal H}'(D,Q)\s]\s|p(t),q(t);\beta\>\,dt \no\\
     &&=\int_0^T\s[\s -q(t)\s {\dot p}(t)-H(p,q)\s]\,dt \;, \en
    where
    \bn H(p,q)\hskip-1.2em&&\equiv H'(p\s q,q)=\<p,q;\beta|\s {\cal H}'(D,Q)\s |p,q;\beta\>\no\\
    &&=\<\beta|\s {\cal H}'(D+p\s qQ,qQ)|\beta\>\no\\
    &&={\cal H}'(p\s q,q)+{\cal O}(\hbar;p,q)\;.\en
 Observe that the quantum function $({\cal H}')$, is the same as the classical function $(H')$, apart from $\hbar$ corrections,
  not unlike what we found in the
 canonical case which led to Cartesian coordinates. However, in the affine case the Fubini-Study metric is
 $ds^2=(\beta\hbar)^{-1}\s q^2\,dp^2+(\beta\hbar)\s q^{-2}\s dq^2$, which describes a Poincare half-plane, a geodetically complete
 space of constant negative curvature $(-2/\beta\hbar)$.  Hence, when $q>0$, the `right phase space coordinates' are {\it not} Cartesian!

\section*{Appendix}
Recall that an unbounded operator, say $A$,  cannot act on all vectors $|\phi\>$ in the Hilbert space $\mathfrak{H}$ but it is restricted to a smaller
domain of vectors, say $D(A)\subset\mathfrak{H}$, which ensures that a vector $|\psi\>\in D(A)$ also satisfies $A|\psi\>\in\mathfrak{H}$.
Any operator, say$A$, has an adjoint $A^\dagger$ which enters expressions such as $\<\psi|A^\dagger|\phi\>=\overline{\<\phi|A|\psi\>};$, where the
over-line denotes complex conjugation. With these forgoing definitions, an operator, say $B$, is Hermitian provided
$
B^\dagger=B$ on $D(B)\subseteq D(B^\dagger)$, while an operator, say $C$, is self adjoint provided $C^\dagger=C$ on
$D(C)=D(C^\dagger)$; thus a self-adjoint operator is also Hermitian, but the converse may be false. This distinction between Hermitian and self
adjoint is critical because it is only self-adjoint operators that serve as generators of unitary operators basically because they only
have real spectra. Specifically, for real labels, $C|c_k\>=c_k|c_k\>$ for which $\<c_j|c_k\>=\delta_{j,k}$ and $C|c\>=c|c\>$ for which
$\<c|c'\>=\delta(c-c')$.
Hence, the operator $C$ generates a spectral realization  of the form $C=\sum_k\s c_k\s \s|c_k\>\<c_k|+\tint_a^b \s c\s|c\>\<c|\,dc$ including,
in general, a sum over discrete vectors $\{|c_k\>\}$ and an integral over continuously labeled vectors $\{|c\>\}$.

\end{document}